\documentclass[conference]{IEEEtran}

\IEEEoverridecommandlockouts
\usepackage{cite}
\usepackage{amsmath,amssymb,amsfonts}
\usepackage{graphicx}
\usepackage{textcomp}
\usepackage{subcaption}
\usepackage[colorlinks=true, linkcolor=black, urlcolor=black, citecolor=black]{hyperref}

\begin{document}

\title{When Audio Separation Hurts Zero-Shot ASR: Evaluating SAM-Audio with Whisper on Bengali and English Speech}

\author{
\IEEEauthorblockN{Akif Islam\IEEEauthorrefmark{1}, Raufun Nahar\IEEEauthorrefmark{2}, Md. Ekramul Hamid\IEEEauthorrefmark{1}}

\IEEEauthorblockA{iamakifislam@gmail.com, nahar@anan-nct.ac.jp, ekram\_hamid@ru.ac.bd}

\IEEEauthorblockA{\IEEEauthorrefmark{1}Department of Computer Science and Engineering, University of Rajshahi, Rajshahi, Bangladesh}

\IEEEauthorblockA{\IEEEauthorrefmark{2}Anan National College of Technology, Tokushima, Japan}
}

\maketitle

\begin{abstract}
Recent advances in automatic speech recognition (ASR) and
speech enhancement have strengthened the common belief that
cleaner audio should lead to more accurate transcription. In
this work, we examine whether this assumption holds for modern
zero-shot ASR systems. We conduct a structured empirical study
of SAM-Audio as a preprocessing step for zero-shot transcription
with OpenAI Whisper. Five Whisper variants are evaluated on
noisy Bengali and English speech datasets. On the English
dataset, SAM-Audio increases the average PSNR from
32.28\,dB to 35.99\,dB and achieves higher PSNR for
71.84\% of the utterances. However, WER and CER increase
in every evaluated model--dataset configuration. On the Bengali
dataset, Whisper \textit{large-v3} WER increases from 65.83\%
to 77.35\%, while CER increases from 24.13\% to 34.74\%. On the English dataset, Whisper \textit{base} WER increases
from 10.53\% to 21.66\%, while CER increases from 4.48\%
to 12.50\%. Utterance-level analysis further shows that the
degradation affects a substantial portion of the evaluated
samples, although its severity varies across Whisper variants.
These findings demonstrate that improved signal-level quality
does not necessarily lead to better zero-shot ASR performance
and that denoising can reduce recognition accuracy.
\end{abstract}


\section{Introduction}
Automatic Speech Recognition (ASR) has seen tremendous advancements over the past few years and has been adopted in many real-life applications including voice assistants, meeting transcription, multimedia indexing, and accessibility technologies\cite{hinton2012deep, chan2016listen,openai2022whisper}. But in the real world, speech isn't always pristine. The audio captured from sources like YouTube, mobile devices, and conversational settings often includes background noise, reverberation, compression effects, and multiple speakers. These distortions are still one of the most challenging factors to achieve a reliable ASR performance and are still a strong driving force to research in noise-robust speech processing\cite{li2014overview}.

One of the most popular and intuitive approaches to tackle this problem is called \emph{speech enhancement} and is usually used as a pre-processing phase prior to the recognition process \cite{loizou2013speech}. The recent developments of deep learning have made enhancement models much more capable, and sometimes the speech they enhanced sounds perceptually clearer to the human listeners\cite{wang2018supervised, fu2021metricganplus}.

Among these models, one is the general audio separation model called \emph{SAM-Audio} which combines multimodal prompting to separate the desired sound sources from complex sound scenes\cite{shi2025samaudio}. SAM-Audio is a flexible and controllable separation system based on a diffusion-transformer architecture and trained on large-scale audio corpora, which goes beyond the classic category-specific enhancement.

Meanwhile, modern end-to-end ASR systems like OpenAI's \emph{Whisper} have proven to be robust in a wide range of acoustic conditions, thanks to training on large-scale weakly-supervised multilingual datasets \cite{openai2022whisper}. Whisper is often employed in a \emph{zero-shot} fashion, meaning that the pre-trained models are applied directly to new datasets without adapting them to the task at hand. A key assumption that is not always explored is: \emph{Does perceptual audio quality improvement with SAM-Audio translate to recognition accuracy improvement?}

Previous studies have investigated the interaction between speech enhancement and ASR, but they have mostly focused on traditional speech enhancement methods or task-aware optimization of speech enhancement\cite{kinoshita2020improving, ochiai2017distortion}. To the best of our knowledge, there is no previous study that systematically explores the effects of SAM-Audio preprocessing on zero-shot ASR performance. This is significant because the enhancement models at the foundation level differ from the traditional denoisers in terms of architecture and training goals, and their relationship with robust end-to-end ASR systems is not well understood.

In this paper, we perform an empirical study to investigate whether SAM-Audio can be used to enhance the zero-shot ASR performance. We test several different versions of Whisper on two language-diverse datasets: a noisy speech dataset in English and a real-world YouTube corpus in Bengali. As one might have expected, our experiments show that in many cases, SAM-Audio actually has a detrimental effect on Word Error Rate (WER) and Character Error Rate (CER), despite the fact that the signal-level quality has been improved. This indicates that there may be a discrepancy between the acoustic properties added by enhancement and the distribution of signals learnt in ASR pretraining.

\subsection{Contributions}

To address this, we conduct the first systematic study of SAM-Audio as a preprocessing step for Whisper-based zero-shot ASR and ask a key question: \emph{Does advanced denoising of perceptual audio improve zero-shot ASR accuracy?} We perform cross-lingual experiments on our collected noisy Bengali YouTube speech data set and a publicly available noisy English speech data set, and demonstrate that by using aggregate metrics and utterance-level analysis, SAM-Audio can significantly improve signal quality while harming recognition accuracy. The results offer empirical evidence of the mismatch between perceptual enhancement of the signal by humans and the robustness of recognition by the machine, and warn against the blind use of powerful denoisers in zero-shot ASR pipelines.

\section{Literature Review}

Speech enhancement has always been considered as a practical approach to enhance the performance of Automatic Speech Recognition (ASR) in noisy conditions. Background noise and reverberation may have a tremendous impact on recognition accuracy, and this has led to decades of research into algorithms that suppress noise while maintaining speech intelligibility. In both early and modern research, enhancement is regarded as a front-end module to powerful ASR systems.

There are several works that report that if the enhancement pipelines are designed properly, then the recognition accuracy can be improved. For example, a single channel time domain denoising method was able to reduce Word Error Rate by over \textbf{30\% relative} on the CHiME-4 benchmark \cite{kinoshita2020improving}. Likewise, combining GAN-based speech enhancement with retraining strategies has led to measurable WER gains \cite{pascual2017segan} and neural acoustic echo cancellation has demonstrated WER gains of up to \textbf{57\%} over signal-processing baselines \cite{howard2021aec}. All of these studies confirm the general opinion that it should be easier to automatically transcribe cleaner speech.

But the connection of enhancement and recognition is not always a good one. Enhancement algorithms can introduce artifacts, modify spectral cues, or mismatch the acoustic distributions used for training of the ASR \cite{ochiai2017distortion,iwamoto2022analyzing,
ochiai2024rethinking}. These distortions can counteract the noise suppression effect, especially for modern end-to-end models which are already robust to noisy data.

This complex relationship is increasingly emphasized in recent studies. For instance, a systematic study of denoising using MetricGAN on various contemporary ASR systems such as Whisper found that better audio does not always lead to better recognition accuracy and in some cases, the original noisy speech was actually better recognized than the denoised version \cite{chondhekar2025when}. These observations suggest that perceptual improvement is not necessarily a good predictor of ASR gains \cite{sato2021should,chondhekar2025when}.

While previous research has shown that speech enhancement can benefit and degrade ASR performance, most of the studies have investigated traditional denoising architectures or enhancement strategies that are optimized along with the recognition model. We are not aware of any previous work that systematically tested SAM-Audio as a preprocessing method for automatic speech recognition. This gap is significant since SAM-Audio is a new class of foundation-scale, multimodal audio models that are not trained specifically for ASR. So, its performance with modern recognition systems is unclear, especially in zero-shot scenarios where models like Whisper are used without adaptation.

In order to explore this under-researched nexus, this work is driven by the central question:

Does SAM-Audio-based separation model improve zero-shot ASR accuracy, or can it degrade recognition despite improving signal-level quality?

\begin{figure}[!t]
    \centering
    \includegraphics[width=0.6\linewidth]{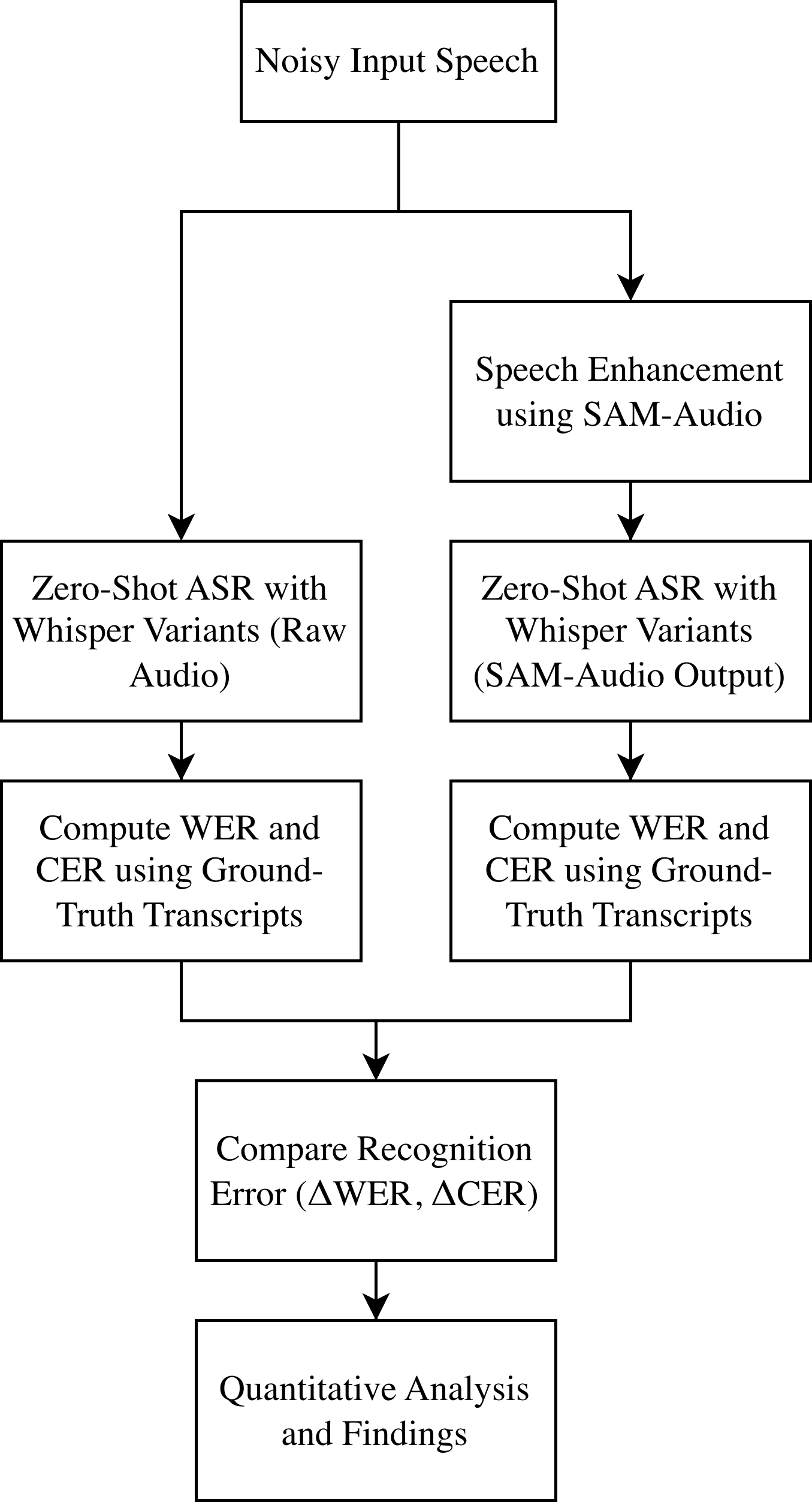}
    \caption{Experimental pipeline for evaluating the impact of SAM-Audio on zero-shot ASR. Noisy speech is transcribed directly and after SAM-Audio enhancement using Whisper variants, and recognition errors (WER/CER) are compared to quantify denoising-induced performance changes.}
    \label{fig:methodology_pipeline}
\end{figure}

\begin{figure*}[!t]
    \centering
    \begin{subfigure}[t]{0.49\textwidth}
        \centering
        \includegraphics[width=\linewidth]{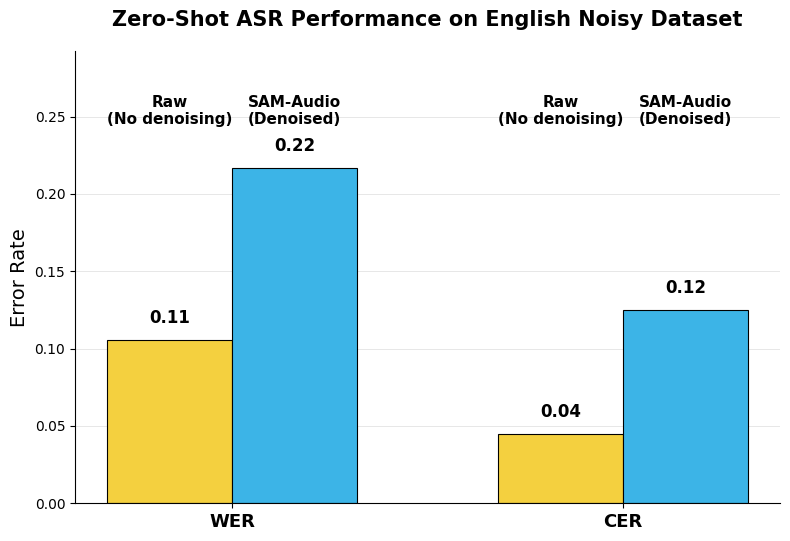}
        \caption{English noisy dataset (Whisper-\textit{base}).}
    \end{subfigure}
    \hfill
    \begin{subfigure}[t]{0.49\textwidth}
        \centering
        \includegraphics[width=\linewidth]{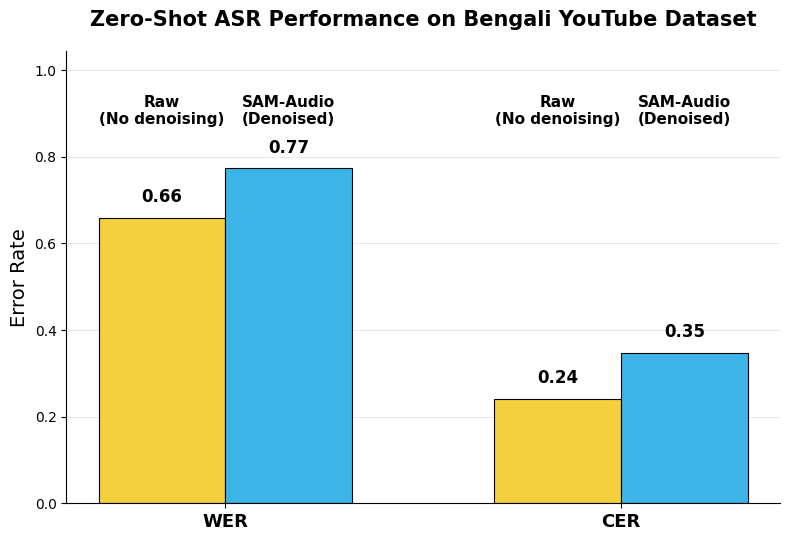}
        \caption{Bengali YouTube dataset (Whisper-\textit{large-v3}).}
    \end{subfigure}
    
    \caption{Impact of SAM-Audio preprocessing on zero-shot ASR performance for the best-performing Whisper models across two noisy speech datasets. Despite improving perceptual signal quality, SAM-Audio denoising consistently results in higher Word Error Rate (WER) and Character Error Rate (CER) compared to raw audio, indicating a mismatch between perceptual enhancement and recognition accuracy.}
    
    \label{fig:sam_audio_comparison}
\end{figure*}

\section{Methodology}

The datasets, the preprocessing pipeline, the ASR models, and the evaluation protocol are described here to investigate the impact of speech denoising on zero-shot ASR performance.

\subsection{Datasets}

To make sure that our findings are not language- or dataset-specific, we perform experiments on two noisy speech datasets that are spoken in two different languages.

\subsubsection{Bangla YouTube Vlogger Dataset (BanglaYTV)}

The first one is a noisy Bengali speech dataset, which we collected from YouTube, mainly from popular Bengali travel vlogging channels. The data set consists of 13.80 hours of audio and is recorded in real-world settings, not in a controlled studio environment. It contains a variety of acoustic problems like background traffic noise, overlapping speech, speech recorded while eating, ambient music, and other distortions typical of user generated content.

In this study, we make use of the test split of the data set to provide a controlled and unbiased comparison of the zero-shot ASR performance on the raw and denoised audio. To achieve high transcription reliability, all test samples are double-annotated by human transcribers and the manually verified ground-truth transcriptions are used to compute recognition error metrics.

\subsection{English Noisy Dataset (MS-SNSD)}

For the English experiments, we use a publicly available noisy
speech dataset named Microsoft Scalable Noisy Speech Dataset (MS-SNSD) \cite{reddy2019scalable}. MS-SNSD is created for training and testing of noise suppression models and contains clean speech recordings along with a variety of environmental noise clips, all of them are single channel 16\,kHz \texttt{.wav} files. A paired evaluation data set is also provided: noisy\_test and clean\_test reference set, which can be used to make objective comparisons between noisy and processed speech.

One of the most important properties of MS-SNSD is that it has a scalable data generation recipe that can be used to mix clean speech and noise at controllable signal-to-noise ratio (SNR) levels to generate noisy speech across a range of acoustic conditions \cite{reddy2019scalable}. The noise collection includes several real world categories (e.g., traffic, babble, appliances, typing, vacuum cleaner) to allow for evaluations under various noise types. For our study, we rely on the given noisy test utterances and clean reference utterances for zero-shot ASR evaluation of both the raw noisy audio and the SAM-Audio denoised audio.

\subsection{ASR Models and Zero-Shot Inference}

We test several different variants of the Whisper automatic
speech recognition model, spanning a wide range of model sizes:
\emph{tiny}, \emph{base}, \emph{small}, \emph{medium}, and
\emph{large-v3}. No fine-tuning or adaptation is performed on
either dataset, and all models are evaluated in a strictly
\emph{zero-shot} setting. Fine-tuning is outside the scope of
this study because our objective is to determine whether passing
audio through a separation model such as SAM-Audio improves
signal quality and zero-shot recognition performance. The
models are therefore used directly to transcribe the input audio,
reflecting a common real-world use case in which Whisper is
applied to previously unseen recordings.

Zero-shot inference is first performed on the \emph{raw noisy
audio} from both datasets to establish the baseline ASR
performance.

\subsection{Denoising of speech using SAM-Audio}

To explore the effect of foundation-scale enhancement on zero-shot ASR, all the test utterances are enhanced using \emph{SAM-Audio} \cite{shi2025samaudio} that can isolate the target sources from complex acoustic mixtures by multimodal cues. The \emph{SAM-Audio Small} variant is only used as an external preprocessing due to hardware limitation and is not jointly trained or adapted with Whisper.

The speech is indicated by a text prompt and only one enhanced speech waveform is produced for each speech utterance as per the official implementation \cite{samaudio_github}. To make sure that there is a controlled comparison between the raw noisy audio and the SAM-Audio processed audio, the inference configuration is selected to avoid span prediction and candidate reranking.

This improved audio is then transcribed using the same variants of Whisper model and decoding options as the original recordings, allowing for a direct comparison of the impact of SAM-Audio preprocessing on zero-shot ASR performance.

\subsection{Evaluation Metrics}

ASR performance is measured in terms of Word Error Rate (WER) and Character Error Rate (CER), where WER and CER are the word and character level mismatch between the transcriptions produced by the model and the ground-truth reference. BLEU-F1 and Real-Time Factor (RTF) were also calculated, but WER and CER are the main metrics since they directly impact on transcription accuracy. We also calculate Peak Signal-to-Noise Ratio (PSNR) on the English dataset to evaluate the signal-level enhancement. Alongside the aggregate measures, we also evaluate utterance-level rolling error trajectories to determine whether SAM-Audio has an impact on all samples in the dataset or only on a few.

\begin{table*}[t]
\centering
\caption{Zero-shot ASR performance of Whisper models on the raw noisy Bengali YouTube test set (mean $\pm$ std).}
\label{tab:bengali_raw}
\begin{tabular}{lccccc}
\hline
Model & WER (mean$\pm$std) & CER (mean$\pm$std) & BLEU-F1 (mean$\pm$std) & RTF & Samples \\
\hline
tiny      & 1.1329 $\pm$ 0.5271 & 1.0631 $\pm$ 0.5021 & 0.0000 $\pm$ 0.0000 & 0.1631 & 929 \\
base      & 1.1282 $\pm$ 0.2839 & 1.0033 $\pm$ 0.2727 & 0.0000 $\pm$ 0.0000 & 0.2249 & 929 \\
small     & 1.2742 $\pm$ 0.8783 & 1.1295 $\pm$ 0.9310 & 0.0000 $\pm$ 0.0000 & 0.9933 & 929 \\
medium    & 1.2050 $\pm$ 0.8314 & 0.9956 $\pm$ 0.5988 & 0.0119 $\pm$ 0.0434 & 2.5450 & 929 \\
large-v3  & 0.6583 $\pm$ 0.2466 & 0.2413 $\pm$ 0.1353 & 0.3892 $\pm$ 0.2182 & 0.4392 & 929 \\
\hline
\end{tabular}
\end{table*}

\begin{table*}[t]
\centering
\caption{Zero-shot ASR performance of Whisper models on the SAM-Audio–denoised Bengali YouTube test set (mean $\pm$ std).}
\label{tab:bengali_sam}
\begin{tabular}{lccccc}
\hline
Model & WER (mean$\pm$std) & CER (mean$\pm$std) & BLEU-F1 (mean$\pm$std) & RTF & Samples \\
\hline
tiny      & 1.1878 $\pm$ 0.7705 & 1.0958 $\pm$ 0.6438 & 0.0001 $\pm$ 0.0036 & 0.1936 & 929 \\
base      & 1.1562 $\pm$ 0.3545 & 1.0090 $\pm$ 0.2364 & 0.0000 $\pm$ 0.0000 & 0.2415 & 929 \\
small     & 1.2925 $\pm$ 1.0691 & 1.1742 $\pm$ 0.9452 & 0.0000 $\pm$ 0.0000 & 1.3305 & 929 \\
medium    & 1.2578 $\pm$ 0.8416 & 1.1409 $\pm$ 1.0100 & 0.0046 $\pm$ 0.0277 & 2.9461 & 929 \\
large-v3  & 0.7735 $\pm$ 0.2565 & 0.3474 $\pm$ 0.2230 & 0.2796 $\pm$ 0.2077 & 0.4814 & 929 \\
\hline
\end{tabular}
\end{table*}

\begin{table*}[t]
\centering
\caption{Zero-shot ASR performance of Whisper models on the raw noisy English dataset.}
\label{tab:english_raw}
\begin{tabular}{lccccc}
\hline
Model & WER (mean$\pm$std) & CER (mean$\pm$std) & BLEU-F1 (mean$\pm$std) & RTF & Samples \\
\hline
tiny      & 0.1612 $\pm$ 0.3619 & 0.0748 $\pm$ 0.2481 & 0.8611 $\pm$ 0.2226 & 0.083 & 824 \\
base      & 0.1053 $\pm$ 0.2095 & 0.0448 $\pm$ 0.1304 & 0.9036 $\pm$ 0.1782 & 0.083 & 824 \\
small     & 1.1014 $\pm$ 1.0421 & 0.9762 $\pm$ 1.0783 & 0.1652 $\pm$ 0.3208 & 4.592 & 824 \\
medium    & 0.1865 $\pm$ 0.4005 & 0.1360 $\pm$ 0.3434 & 0.8383 $\pm$ 0.3258 & 2.402 & 824 \\
large-v3  & 1.1895 $\pm$ 0.5057 & 1.1387 $\pm$ 0.5670 & 0.0056 $\pm$ 0.0665 & 6.321 & 824 \\
\hline
\end{tabular}
\end{table*}

\begin{table*}[t]
\centering
\caption{Zero-shot ASR performance of Whisper models on the SAM-Audio–denoised English noisy dataset (mean $\pm$ std).}
\label{tab:english_sam}
\begin{tabular}{lccccc}
\hline
Model & WER (mean$\pm$std) & CER (mean$\pm$std) & BLEU-F1 (mean$\pm$std) & RTF & Samples \\
\hline
tiny      & 0.2663 $\pm$ 0.4383 & 0.1618 $\pm$ 0.3462 & 0.7716 $\pm$ 0.2992 & 0.089 & 824 \\
base      & 0.2166 $\pm$ 0.3366 & 0.1250 $\pm$ 0.2605 & 0.8070 $\pm$ 0.2791 & 0.121 & 824 \\
small     & 1.2120 $\pm$ 1.3916 & 1.0391 $\pm$ 1.1107 & 0.1092 $\pm$ 0.2504 & 4.610 & 824 \\
medium    & 0.2743 $\pm$ 0.5013 & 0.2134 $\pm$ 0.4395 & 0.7681 $\pm$ 0.3671 & 2.971 & 824 \\
large-v3  & 1.2278 $\pm$ 0.6178 & 1.1691 $\pm$ 0.6675 & 0.0017 $\pm$ 0.0355 & 5.942 & 824 \\
\hline
\end{tabular}
\end{table*}

\section{Results and Discussions}
\label{sec:results}

In this section, Zero-shot ASR performance of Whisper models is reported on the BanglaYTV and MS-SNSD dataset. We use Word Error Rate (WER) and Character Error Rate (CER) as the primary metrics, and BLEU-F1 and Real-Time Factor (RTF) are included for completeness (Tables~\ref{tab:bengali_raw}--\ref{tab:english_sam}).

We see this consistent and surprising pattern across all experiments: \textbf{raw noisy audio gives lower recognition error than SAM-Audio--processed audio and this is independent of model size and language}. This contradicts the conventional wisdom that perceptual improvement is always beneficial for ASR in a zero-shot scenario.

\subsection{Zero-Shot ASR on BanglaYTV dataset}
\label{subsec:results_bengali}

Table~\ref{tab:bengali_raw} summarizes the results on the raw
Bengali test set. Among the evaluated Whisper variants,
\textit{large-v3} achieves the lowest WER and CER. However,
performance does not improve monotonically with model size,
as the remaining variants show substantial variation in
recognition error.

Post-SAM-Audio preprocessing (Table~\ref{tab:bengali_sam}) leads to a drop in performance for \emph{all} Whisper variants. For example, \textit{large-v3} worsens from 0.6583 to 0.7735 WER and from 0.2413 to 0.3474 CER. The same trend is observed for smaller models, suggesting it is a systematic effect and not just a failure of one model.

\begin{figure}[t]
\centering
\includegraphics[width=\linewidth]{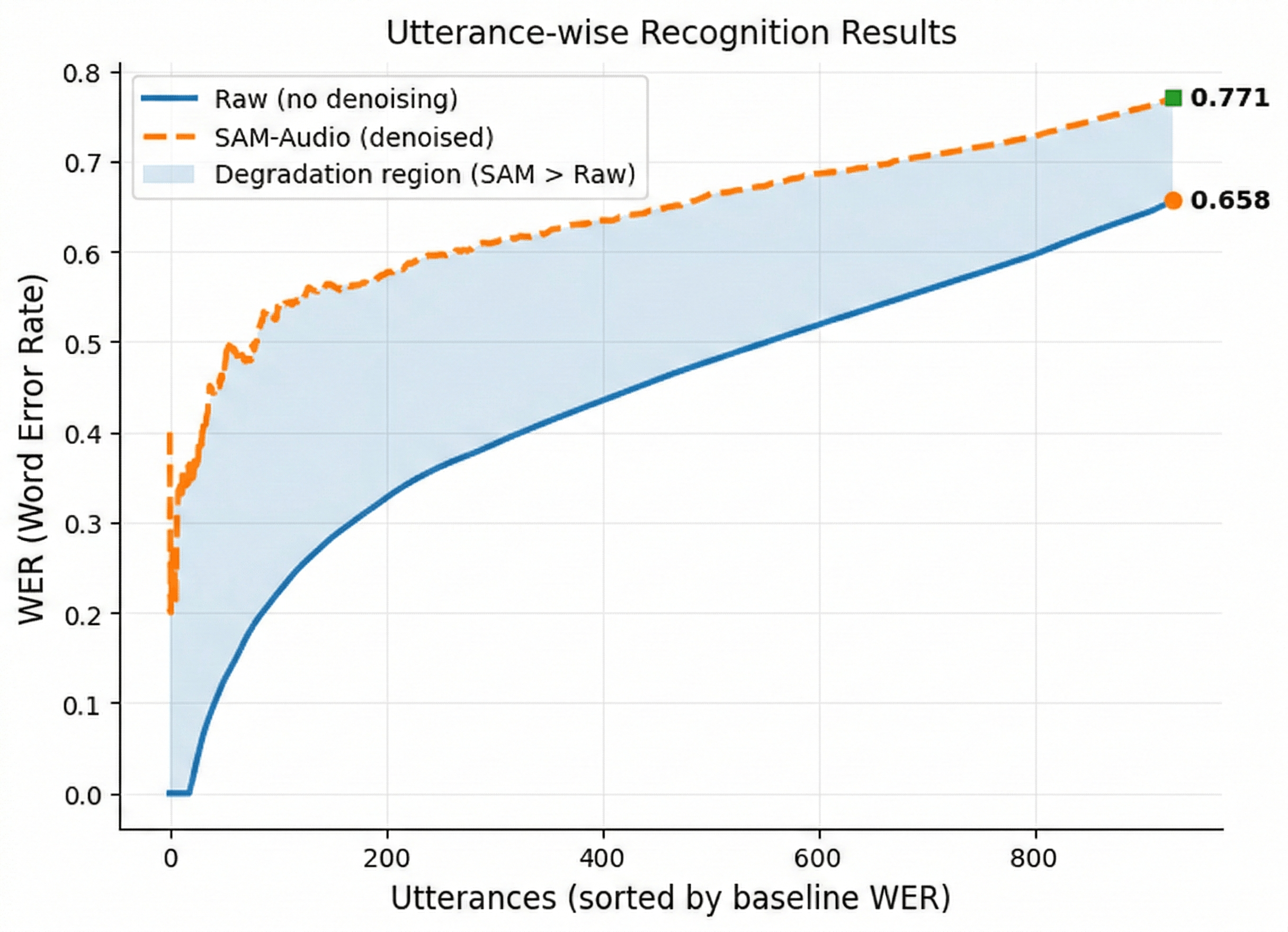}
\caption{Running average WER across Bengali utterances sorted by baseline (raw) WER. The shaded region highlights where SAM-Audio yields higher error than raw audio.}
\label{fig:bengali_running}
\end{figure}

To check if degradation is due to a small set of hard clips, we plot the running average WER after sorting the utterances by their baseline (raw) WER. As seen in Figure~\ref{fig:bengali_running}, the denoised trajectory is generally above the raw baseline for the majority of utterances, indicating a shift in distribution, and not individual outliers.

\subsection{Zero-Shot ASR on English Noisy Dataset}
\label{subsec:results_english}

The same protocol is repeated on an English noisy dataset (MS-SNSD), to examine if the effect extends to other languages. On raw audio (Table~\ref{tab:english_raw}), the \textit{base} model provides the lowest mean WER/CER and larger variants are unstable with this test condition.

Recognition accuracy decreases for all evaluated models after
SAM-Audio preprocessing, as shown in
Table~\ref{tab:english_sam}. The \textit{small} and
\textit{large-v3} variants exhibit particularly high WER values,
which increase from 1.1014 to 1.2120 and from 1.1895 to
1.2278, respectively. The consistent increase across all
variants suggests that the observed degradation is not limited
to the Bengali dataset.

\begin{figure}[!t]
\centering
\includegraphics[width=\linewidth]
{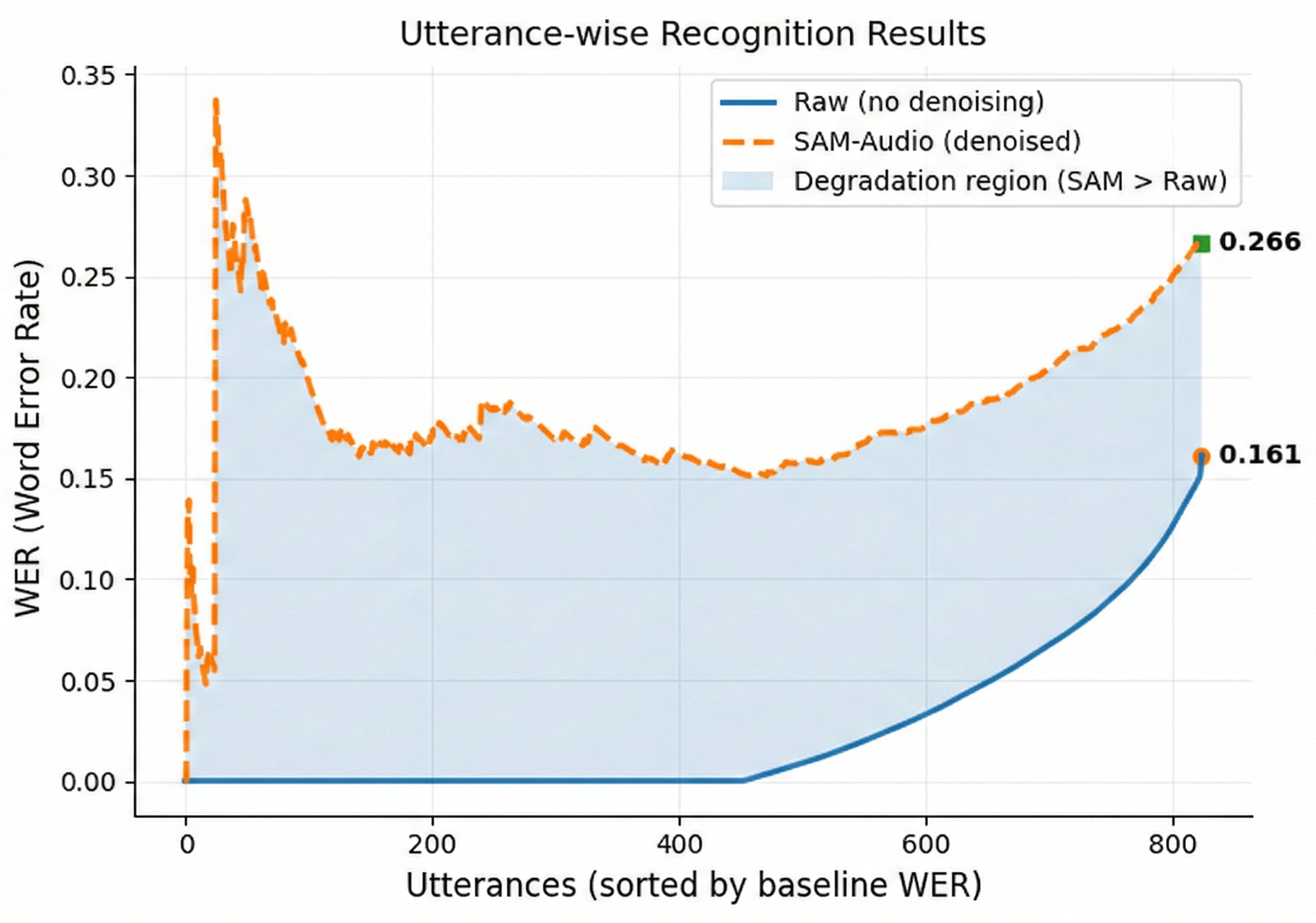}
\caption{Running average WER on English utterances ranked by
baseline raw WER.}
\label{fig:english_running}
\end{figure}

The evidence is at the utterance level, with the running average WER again revealing a persistent difference between the raw and denoised trajectories, as depicted in Figure~\ref{fig:english_running}.

\subsection{Signal Level Quality Analysis using PSNR}
\label{subsec:psnr_results}

For our BanglaYTV dataset, clean reference audio was not
available. However, for the MS-SNSD dataset, both clean and
noisy versions of the audio were available. Therefore, we
performed Peak Signal-to-Noise Ratio (PSNR) analysis on the
English MS-SNSD dataset to check whether the signal-level
quality is improved by SAM-Audio.

PSNR is calculated between the clean audio and the original
noisy audio, and between the clean audio and the
SAM-Audio--denoised audio. A total of 824 matched clean,
noisy, and enhanced utterances were evaluated.

As seen in Table~\ref{tab:psnr_results}, SAM-Audio boosts the
average PSNR from 32.28\,dB to 35.99\,dB, while the PSNR of
71.84\% of the utterances is improved. However, the ASR
results in Tables~\ref{tab:english_raw} and
\ref{tab:english_sam} show that recognition accuracy still
decreases after denoising. This means that zero-shot ASR
performance may not always improve when signal-level quality
improves. PSNR is not reported for the BanglaYTV dataset
because clean reference audio was not available.

\begin{table}[t]
\centering
\caption{PSNR comparison on English noisy data set (dB).}
\label{tab:psnr_results}
\begin{tabular}{lc}
\hline
Condition & PSNR (dB) \\
\hline
Clean vs.\ Noisy        & 32.28 \\
Clean vs.\ SAM-Denoised & \textbf{35.99} \\
\hline
\end{tabular}
\end{table}

\subsection{Why Denoising hurts Zero-Shot ASR?}
\label{subsec:why_denoising_hurts}

This regular degradation following SAM-Audio preprocessing naturally brings up the question: why does perceptually improved audio performance negatively impact recognition in a zero shot ASR setting?

As can be seen from Figures~\ref{fig:bengali_running} and~\ref{fig:english_running}, the running average WER for the SAM-Audio processed speech is consistently higher than the WER for the original speech for most of the utterances in both the datasets. This indicates that the degradation is broadly distributed across the evaluated utterances rather than being caused only by a few difficult clips or outliers. The pattern is consistent with an enhancement-induced input mismatch, although the current experiments do not directly establish the underlying causal mechanism.

Although SAM-Audio increases average signal-level PSNR, the processed signals do not necessarily preserve all acoustic characteristics on which zero-shot ASR models depend. PSNR is used here as a signal-level fidelity measure and does not, by itself, establish improved human-perceived speech quality. Modern ASR systems like Whisper are pretrained on vast amounts of weakly-supervised, noisy audio. If a separation-based denoiser is too aggressive in suppressing or reshaping these components, the improved signal will not necessarily have the same statistical characteristics as the data observed during pretraining, leading to a \emph{distribution mismatch} \cite{sato2021should,iwamoto2022analyzing}.

Also, source separation and enhancement can introduce subtle
distortions and processing artifacts that can affect ASR
performance
\cite{iwamoto2022analyzing,ochiai2024rethinking}. These artefacts are usually not audible to humans, but can interfere with the fine-grained time--frequency patterns used by neural ASR models for decoding. The magnitude of degradation varies across Whisper variants, but the present results do not demonstrate a consistent relationship between model size and sensitivity to enhancement.

In sum, the results indicate a basic difference between \emph{listening quality} and \emph{recognition quality}. From our results, we show that using powerful, foundation-scale denoising models blindly, as a pre-processing step, can actually harm the recognition performance and hence the need for ASR-aware evaluation and integration of speech enhancement methods.

\subsection{Limitations}
\label{sec:limitations}

This study assesses SAM-Audio as an \emph{external preprocessing module} for zero-shot ASR in realistic resource-constrained settings. First, all experiments use the \emph{SAM-Audio Small} variant. We were unable to evaluate the Medium or Large models or more compute-intensive options such as multi-candidate reranking and span prediction. Future work should assess whether these settings change the observed ASR behavior.

Second, we analyze two moderately sized noisy speech datasets. Evaluation on larger corpora and other domains, such as conversational speech or far-field recordings, could provide further insights. Lastly, we consider only a zero-shot ASR setting. Future work could investigate whether joint adaptation or lightweight fine-tuning reduces the observed performance degradation.

Nevertheless, the consistency of the results across languages, model sizes, and utterance-level analyses suggests a mismatch between separation-based enhancement and zero-shot ASR.

\section{Conclusion}

This paper proposes a systematic study on speech denoising for zero-shot ASR. We find a consistent and surprising pattern: across SAM-Audio preprocessing and multiple variants of Whisper, and across noisy Bengali and English speech, signal-level quality is improved but recognition accuracy is often degraded in zero-shot settings. The magnitude of degradation varies across Whisper variants, but the results do not demonstrate a consistent relationship between model size and sensitivity to enhancement. The results contradict the widely held belief that higher audio quality always leads to better ASR performance, thus revealing a gap between signal-level quality and recognition performance. From our results, we show that powerful denoising models can be detrimental to modern ASR systems that already encode significant noise robustness. Due to computational constraints, this study is limited to the SAM-Audio Small variant and zero-shot inference. Future research will evaluate larger SAM-Audio models, recommended inference settings, larger datasets, and joint or adaptive enhancement--recognition approaches to better understand when and how denoising can be helpful for ASR.

\bibliographystyle{IEEEtran}
\bibliography{references}

\end{document}